\documentclass{article}
\usepackage[a4paper]{geometry}
\usepackage{graphicx}
\usepackage[utf8x]{inputenc}
\usepackage{hyperref}
\usepackage{todonotes}
\usepackage{amsmath}
\usepackage{amssymb}
\usepackage{upgreek}
\usepackage{fixmath}
\usepackage{lmodern}
\usepackage{subfig}
\usepackage[sort&compress]{natbib}
\bibpunct{[}{]}{,}{n}{}{}

\renewcommand{\epsilon}{\varepsilon}

\newcommand{\grad}{\nabla}

\newcommand{\D}[2]{\ensuremath{\frac{\partial#2}{\partial#1}}}
\renewcommand{\vec}[1]{\ensuremath{\boldsymbol{#1}}} 

\newcommand{\figplacing}{hbt}
\newcommand{\figwidth}{0.75\linewidth}

\newcommand{\Note}[1]{\ensuremath{\mathrm{^#1}}}

\makeatletter
 \providecommand\phantomcaption{\caption@refstepcounter\@captype}
\makeatother

\title{Particle transport in density gradient driven TE~mode turbulence}
\author{A. Skyman\Note{1}, H. Nordman\Note{1}, P. I. Strand\Note{1}\\
\small \Note{1}Euratom--VR Association, Department of Earth and Space Sciences, \\\small Chalmers University of Technology, SE-412 96 Göteborg, Sweden}
\date{}

\begin{document}
%


\maketitle
\begin{abstract}
\noindent The turbulent transport of main ion and trace impurities in a tokamak device in the presence of steep electron density gradients has been studied.
The parameters are chosen for trapped electron (TE) mode turbulence, driven primarily by steep electron density gradients relevant to H-mode physics, but with a transition to temperature gradient driven turbulence as the density gradient flattens~\cite{Ernst2009}.
Results obtained through non-linear (NL) and quasilinear (QL) gyrokinetic simulations using the GENE code~\cite{Merz2008a, GENE} are compared with results obtained from a fluid model.
Main ion and impurity transport is studied by examining the balance of convective and diffusive transport, as quantified by the density gradient corresponding to zero particle flux (peaking factor).
Scalings are obtained for the impurity peaking with the background electron density gradient and the impurity charge number.
It is shown that the impurity peaking factor is weakly dependent on impurity charge and significantly smaller than the driving electron density gradient.
\end{abstract}

\noindent The compatibility between a reactor-grade plasma and the material walls surrounding the plasma is one of the main challenges facing a magnetic fusion device.
The presence of very low levels of high $Z$ impurities in the core plasma may lead to unacceptable levels of radiation losses and fuel dilution.
Also low $Z$ impurities, in the form of Beryllium or Helium-ash, may result in fuel dilution that severely limits the attainable fusion power~\cite{Harte2010}.
Consequently, the transport properties of impurities is a high priority issue in present experimental and theoretical fusion plasma research.
This is emphasised by the the new ITER-like wall experiment in JET~\cite{Matthews2009}, where a Beryllium-clad first wall in the main chamber, combined with carbon and tungsten tiles in the divertor, will be tested for the first time.
 
The transport of main fuel as well as impurities in the core region of tokamaks is expected to be dominated by turbulence driven by Ion Temperature Gradient (ITG) modes and Trapped Electron (TE) modes.
The main drives for the ITG/TE~mode instabilities are gradients of temperature and density combined with unfavourable magnetic curvature.
Most of the theoretical studies of turbulent particle transport have been devoted to temperature gradient driven ITG~and TE~modes, using both fluid, quasilinear (QL) and nonlinear (NL) gyrokinetic models~\cite{Frojdh1992, Basu2003, Estrada-Mila2005, Naulin2005, Priego2005, Fulop2006, Bourdelle2007, Dubuit2007, Camenen2009, Fulop2010, Futatani2010, Hein2010, Moradi2010, Fulop2011, Nordman2008, Angioni2006, Nordman2007a, Angioni2007, Angioni2009a, Fulop2009, Nordman2011, Skyman2011a}.
Much less effort has been devoted to particle transport in regions with steep density gradients. 
The density gradient provides a drive for TE~modes which may dominate the temperature gradient drive for plasma profiles with $R/L_{n_e}>R/L_{T_e}$.
This may occur in connection with the formation of transport barriers, like the high confinement mode edge pedestal, in fusion plasmas.

In the present letter, the turbulent transport of main ion and trace impurities in tokamaks is investigated through nonlinear (NL) gyrokinetic simulations using the GENE code.
The main part considers collisionless TE~modes driven by density gradients but with a transition to temperature gradient driven TE~modes as the density profiles flattens.
The impurity density gradient for zero impurity flux is calculated for varying background electron density gradient drive and for a range of impurity species.
This study complements recent studies~\cite{Nordman2011, Skyman2011a}
on temperature gradient driven TE~and ITG~mode impurity transport. 
The results are compared with QL kinetic simulations and a computationally efficient multi fluid model, suitable for use in predictive transport simulations.
Of particular interest is the sign of the impurity convective flux and the degree of impurity peaking in the presence of strong background electron density gradients.


The models used have been described in detail elsewhere, see~\cite{Nordman2011} and references therein, only a brief summary is given here.
The NL and QL GENE simulations were performed in a flux tube geometry, in a low $\beta$ ($\beta=10^{-4}$) s--$\alpha$ equilibrium~\cite{Jenko2000, Dannert2005a, Dannert2005b, Merz2008a}.

In order to ensure that the resolution was adequate, the resolution was varied separately for the perpendicular, parallel and velocity space coordinates, and the effects of this on the mode structure, $k_\perp$ spectra and flux levels were investigated. 
The resolution was then set sufficiently high for the effects on these indicators to have converged.
For a typical NL simulation for main ions, fully kinetic electrons, and one trace species, a resolution of $n_{x} \times n_{y} \times n_{z} = 96 \times 96 \times 24$ grid points in real space and of $n_{v} \times n_{\mu} = 48 \times 12$ in velocity space was chosen.
For QL~GENE simulations the box size was set to $n_{x} \times n_{y} \times n_{z} = 8 \times 1 \times 24$  and $n_{v} \times n_{\mu} = 64 \times 12$ respectively.
The impurities were included self-consistently as a third species in the simulations, with the trace impurity particle density $n_Z/n_e = 10^{-6}$ in order to ensure that they have a negligible effect on the turbulence.

For the fluid simulations, the Weiland multi-fluid model~\cite{Weiland2000} is used to derive the main ion, impurity, and trapped electron density response from the corresponding fluid equations in the collisionless and electrostatic limit.
The free electrons are assumed to be Boltzmann distributed.
The equations are closed by the assumption of quasineutrality.
An eigenvalue equation for TE~and ITG~modes is obtained in the presence of impurities.
The eigenvalue equation is solved for general mode width~\cite{Weiland2000}.
Alternatively, a strongly ballooning eigenfunction with $k_\parallel^2=\left(3 q^2 R^2\right)^{-1}$ can be used for magnetic shear $s \sim 1$~\cite{Hirose1994}.
The eigenvalue equation is then reduced to a system of algebraic equations that is solved numerically.

\label{sec:PF}
The main ion and impurity particle fluxes can then be written as:

\begin{equation} \label{eq:Gamma_derivation}
\Gamma_{j} =
\left<\delta n_j v_{\vec{E}\times\vec{B}}\right> = 
-n_j \rho_s c_s\left<\widetilde{n}_j\frac{1}{r}\D{\theta}{\widetilde{\phi}}
\right>.
\end{equation}

\noindent The angled brackets imply a time and space average over all unstable modes. Performing this averaging for a fixed length scale $k_\theta\rho_s$ of the turbulence, the particle flux can be written:


\begin{equation}
 \label{eq:transport}
 \frac{R\Gamma_j}{n_j} = D_j\frac{R}{L_{n_j}} + D_{T_j}\frac{R}{L_{T_j}} + R V_{p,j}.
\end{equation}


The first term in equation~\eqref{eq:transport} corresponds to diffusion, the second to the thermodiffusion and the third to the convective velocity (pinch), where $1/L_{n_j}=-\grad n_j/n_j$, $n_j$ is the density of species $j$ and $R$ is the major radius of the tokamak
The pinch contains contributions from curvature and parallel compression effects.
These have been described in detail in previous work~\cite{Nordman2007a, Nordman2008, Nordman2011, Angioni2006}.
For trace impurities, equation~\eqref{eq:transport} can be uniquely written $R\Gamma_Z/n_Z =D_Z R/L_{n_Z} + RV_Z$, where $D_Z$ is the impurity diffusion coefficient and $V_Z$ is the impurity convective velocity.

The zero-flux impurity density gradient (peaking factor) is defined as $PF_Z=-RV_Z/D_Z$ for the value of the impurity density gradient that gives zero impurity flux.
Solving the linearised equation~\eqref{eq:transport} for $R/L_{n_Z}$ with $\Gamma_Z = 0$ yields the interpretation of $PF_Z$ as the gradient of zero impurity flux, and it quantifies the balance between convective and diffusive impurity transport.

The main parameters used in the simulations are summarised in table~\ref{tab:parameters}.
The parameters where chosen to represent an arbitrary tokamak geometry at about mid radius, and do not represent any one particular experiment.
A a moderately steep electron temperature gradient ($R/L_{T_e}=5.0$) together with a flatter ion temperature gradient ($R/L_{T_i, Z}=2.0$) were used to promote TE~mode dominated dynamics.
Following~\cite{Ernst2009}, the background density gradient for the base scenario was set higher than the temperature gradient, to ensure density gradient driven dynamics.
In order to preserve quasineutrality $\grad n_e=\grad n_i$ was used. 

First, the main ion particle flux ($\Gamma_p$) is studied.
Time averaged fluxes are calculated from time series of NL~GENE data after convergence, as illustrated in figure~\ref{fig:time_series}.
The scalings of $\Gamma_p$ with the electron density gradient obtained from NL~GENE and fluid simulations are shown in figure~\ref{fig:Gamma_p}.
The large transport found in NL~GENE simulations is an indication of the stiffness of the gyrokinetic model, and is often seen in fixed-gradient simulations of turbulence.
The fluid model shows a similar scaling of the main ion flux, but the transport is smaller for $R/L_{n_e} > 3.0$.
The main ion density gradient corresponding to zero ion flux ($PF_p$) can be found by similar means to that of $PF_Z$, however, since the trace approximation is not valid for the main ions, the zero-flux gradient has to be found explicitly by varying $\grad n_p$ until the condition $\Gamma_p=0$ is satisfied.
The NL~GENE results presented in figure~\ref{fig:Gamma_p} indicate that for the present parameters, lower density gradients only result in $\Gamma_p \rightarrow 0$.
The the fluid model gives a small outward flux in the limit $R/L_{n_e} \rightarrow 0$.
Neither model results in flux reversal for TE~mode driven turbulence.

Next, the scaling of the impurity transport with the background density gradient ($R/L_{n_e}$) is investigated.
The results for the impurity peaking factor are shown in figure~\ref{fig:omn_TEM}.
We note that the impurity peaking saturates with $PF_Z\approx 2.0$ for large values of the electron density gradient.
The QL results tend to consistently overestimate the peaking factors, while the fluid model gives results that are somewhat below the NL GENE results for the steeper gradients.
The fluid results show a considerably less dramatic dependency of the peaking factor than the gyrokinetic results, both of which show a strong decrease in $PF_Z$ as the electron density profiles flatten.
This is observed for all values of the impurity charge number.
As the background density profile becomes more peaked, a corresponding increase in impurity transport is expected.
This is illustrated in figure~\ref{fig:D_RV}, where scalings, obtained from NL~GENE simulations, of the diffusivity ($D_Z$) and convective velocity ($R V_Z$) with $R/L_{n_e}$ are shown.
Although $D_Z$ and $R V_Z$ strongly increase with $R/L_{n_e}$, the impurity peaking ($PF_Z=-RV_Z/D_Z$) is only weakly sensitive to the electron density gradient.
For $R/L_{n_e}\lesssim 2.0$ the impurity peaking factor is not well defined, since both $D_Z$ and $R V_Z$ go to zero.
 The corresponding linear eigenvalues are displayed in figure~\ref{fig:omn_TEM_eigens}.
The fluid and gyrokinetic results are in qualitative agreement, showing an growthrate that increases uniformly with $R/L_{n_e}$.
The results indicate a smooth transition from density gradient driven to temperature gradient driven TE~mode turbulence, which dominates for $R/L_{n_e} \lesssim R/L_{T_e}$~\cite{Ernst2009}.

The scaling of the impurity peaking factor with impurity charge ($Z$), with $R/L_{n_e}$ as a parameter, is illustrated in figure~\ref{fig:Z}.
The models show only a very weak scaling, with $PF_Z$ falling toward saturation for higher $Z$.
The results are similar to those for the temperature gradient driven TE~mode reported in~\cite{Skyman2011a}.
Notably, the QL~GENE simulations overestimate the peaking factors, whereas the fluid results are lower than the peaking factors obtained from NL~GENE simulations.
The trend observed for low $Z$ impurities is reversed compared to trends reported in e.g.~\cite{Nordman2011} for ITG~mode driven impurity transport.
The qualitative difference can be understood from the $Z$-dependent thermodiffusion in equation~\eqref{eq:transport}, which is outward for ITG~modes and inward for TE~modes.

%
%

In summary, the turbulent transport of main ion and trace impurities in regions of steep density gradients has been investigated through nonlinear (NL) gyrokinetic simulations using the GENE code.
The main part has considered collisionless TE~modes driven by density gradients but with a transition to temperature gradient driven TE~modes as the density profiles flattens.
The results for the impurity density gradient of zero particle flux (peaking factor) have been compared with QL kinetic simulations and a reduced and computationally efficient multi-fluid model, suitable for use in predictive transport simulations.

For the parameters studied, qualitative agreement between gyrokinetic and fluid results has been obtained for the scaling of the impurity peaking factor with both the background density gradient and the impurity charge.
In the region of steep electron density gradients, it was shown that the impurity peaking factor saturates at values significantly smaller than the driving electron density gradient.
It was noted that for the chosen length scales, the QL~GENE results generally overestimate the peaking factor, whereas the fluid results are close to or lower than the NL~GENE results.
The scaling of the peaking factor with impurity charge was observed to be weak, with a slight increase in the impurity peaking factor observed in the gyrokinetic results for low impurity charge numbers.


\section*{Acknowledgements}
The simulations were performed on resources provided on the Lindgren~\cite{Lindgren} and HPC-FF~\cite{HPC-FF} high performance computers, by the
Swedish National Infrastructure for Computing (SNIC) at Paralleldatorcentrum
(PDC) and the European Fusion Development Agreement (EFDA), respectively.

The authors would like to thank Frank Jenko, Tobias G\"orler, M.~J. P\"uschel, and the rest of the GENE~team~\cite{GENE} at IPP--Garching for their help with the gyrokinetic simulations.

\bibliographystyle{unsrtnat}
\bibliography{fusion.bib}

\begin{thebibliography}{34}
\providecommand{\natexlab}[1]{#1}
\providecommand{\url}[1]{\texttt{#1}}
\expandafter\ifx\csname urlstyle\endcsname\relax
  \providecommand{\doi}[1]{doi: #1}\else
  \providecommand{\doi}{doi: \begingroup \urlstyle{rm}\Url}\fi

\bibitem[Ernst et~al.(2009)Ernst, Lang, Nevins, Hoffman, and Chen]{Ernst2009}
D.~R. Ernst, J.~Lang, W.~M. Nevins, M.~Hoffman, and Y.~Chen.
\newblock Role of zonal flow in trapped electron mode turbulence through
  nonlinear gyrokinetic particle and continuum simulation.
\newblock \emph{Phys. Plasmas}, 16\penalty0 (5):\penalty0 055906, 2009.

\bibitem[Merz(2008)]{Merz2008a}
F.~Merz.
\newblock \emph{Gyrokinetic Simulation of Multimode Plasma Turbulence}.
\newblock Ph.d. thesis (monography), Westf\"alischen Wilhelms-Universit\"at
  M\"unster, 2008.

\bibitem[GEN()]{GENE}
The {GENE} code.
\newblock URL \url{http://www.ipp.mpg.de/~fsj/gene/}.

\bibitem[Harte et~al.(2010)Harte, Suzuki, Kato, Sakaue, Kato, TAmura, Sudo,
  D'Arcy, Sokell, White, and O'Sullivan]{Harte2010}
C.~S. Harte, C.~Suzuki, T.~Kato, H.~A. Sakaue, D.~Kato, N.~TAmura, S.~Sudo,
  R.~D'Arcy, E.~Sokell, J.~White, and G.~O'Sullivan.
\newblock Tungsten spectra recorded at the {LHD} and comparison with
  calculations.
\newblock \emph{J. Phys. B}, 43\penalty0 (20):\penalty0 205004, 2010.

\bibitem[Matthews et~al.(2009)Matthews, Edwards, Greuner, Loving, Maier,
  Martens, Philipps, Riccardo, Rubel, Ruset, Scmidt, and
  Vllediew]{Matthews2009}
G.~F. Matthews, P.~Edwards, H.~Greuner, A.~Loving, H.~Maier, Ph. Martens,
  V.~Philipps, V.~Riccardo, M.~Rubel, C.~Ruset, A.~Scmidt, and E.~Vllediew.
\newblock Current status of the {JET ITER-like Wall Project}.
\newblock \emph{Phys. Scr.}, 2009\penalty0 (T138):\penalty0 014030, 2009.

\bibitem[Fr\"ojdh et~al.(1992)Fr\"ojdh, Liljestr\"om, and Nordman]{Frojdh1992}
M.~Fr\"ojdh, M.~Liljestr\"om, and H.~Nordman.
\newblock Impurity effects on $\eta_i$ mode stability and transport.
\newblock \emph{Nucl. Fusion}, 32\penalty0 (3):\penalty0 419, 1992.

\bibitem[Basu et~al.(2003)Basu, Jessen, Naulin, and Rasmussen]{Basu2003}
R.~Basu, T.~Jessen, V.~Naulin, and J.~Juul Rasmussen.
\newblock Turbulent flux and the diffusion of passive tracers in electrostatic
  turbulence.
\newblock \emph{Phys. Plasmas}, 10\penalty0 (7):\penalty0 2696, 2003.

\bibitem[Estrada-Mila et~al.(2005)Estrada-Mila, Candy, and
  Waltz]{Estrada-Mila2005}
C.~Estrada-Mila, J.~Candy, and R.W. Waltz.
\newblock Gyrokinetic simulations of ion and impurity transport.
\newblock \emph{Phys. Plasmas}, 12\penalty0 (2):\penalty0 022305, 2005.

\bibitem[Naulin(2005)]{Naulin2005}
V.~Naulin.
\newblock Impurity and trace tritium transport in tokamak edge turbulence.
\newblock \emph{Phys. Rev. E}, 71\penalty0 (1):\penalty0 015402, 2005.
\newblock \doi{10.1103/PhysRevE.71.015402}.

\bibitem[Priego et~al.(2005)Priego, Garcia, Naulin, and Rasmussen]{Priego2005}
M.~Priego, O.~E. Garcia, V.~Naulin, and J.~Juul Rasmussen.
\newblock Anomalous diffusion, clustering, and pinch of impurities in plasma
  edge turbulence.
\newblock \emph{Phys. Plasmas}, 12\penalty0 (6):\penalty0 062312, 2005.

\bibitem[F\"ul\"op and Weiland(2006)]{Fulop2006}
T.~F\"ul\"op and J.~Weiland.
\newblock Impurity transport in {ITER}-like plasmas.
\newblock \emph{Phys. Plasmas}, 13\penalty0 (11):\penalty0 112504, 2006.

\bibitem[Bourdelle et~al.(2007)Bourdelle, Garbet, Imbeaux, Casati, Dubuit,
  Guirlet, and Parisot]{Bourdelle2007}
C.~Bourdelle, X.~Garbet, F.~Imbeaux, A.~Casati, N.~Dubuit, R.~Guirlet, and
  T.~Parisot.
\newblock A new gyrokinetic quasilinear transport model applied to particle
  transport in tokamak plasmas.
\newblock \emph{Phys. Plasmas}, 14\penalty0 (11):\penalty0 112501, 2007.

\bibitem[Dubuit et~al.(2007)Dubuit, Garbet, Parisot, Guirlet, and
  Bourdelle]{Dubuit2007}
N.~Dubuit, X.~Garbet, T.~Parisot, R.~Guirlet, and C.~Bourdelle.
\newblock Fluid simulations of turbulent impurity transport.
\newblock \emph{Phys. Plasmas}, 14\penalty0 (4):\penalty0 042301, 2007.

\bibitem[Camenen et~al.(2009)Camenen, Peeters, Angioni, Casson, Hornsby,
  Snodin, and Strintzi]{Camenen2009}
Y.~Camenen, A.~G. Peeters, C.~Angioni, F.~J Casson, W.~A Hornsby, A.~P. Snodin,
  and D.~Strintzi.
\newblock Impact of the background toroidal rotation on particle and heat
  turbulent transport in tokamak plasmas.
\newblock \emph{Phys. Plasmas}, 16\penalty0 (1):\penalty0 012503, 2009.

\bibitem[F\"ul\"op et~al.(2010)F\"ul\"op, Braun, and Pusztai]{Fulop2010}
T.~F\"ul\"op, S.~Braun, and I.~Pusztai.
\newblock Impurity transport driven by ion temperature gradient turbulence in
  tokamak plasmas.
\newblock \emph{Phys. Plasmas}, 17\penalty0 (6):\penalty0 062501, 2010.

\bibitem[Futatani et~al.(2010)Futatani, Garbet, Benkadda, and
  Dubuit]{Futatani2010}
S.~Futatani, X.~Garbet, S.~Benkadda, and N.~Dubuit.
\newblock Reversal of impurity pinch velocity in tokamaks plasma with a
  reversed magnetic shear configuration.
\newblock \emph{Phys. Rev. Lett.}, 104\penalty0 (1):\penalty0 015003, 2010.

\bibitem[Hein and Angioni(2010)]{Hein2010}
T.~Hein and C.~Angioni.
\newblock Electromagnetic effects on trace impurity transport in tokamak
  plasmas.
\newblock \emph{Phys. Plasmas}, 17\penalty0 (1):\penalty0 012307, 2010.

\bibitem[Moradi et~al.(2010)Moradi, Tokar, and Weyssow]{Moradi2010}
S.~Moradi, M.~Z. Tokar, and B.~Weyssow.
\newblock Modeling of impurity effect on drift instabilities in plasmas with
  many ion species.
\newblock \emph{Phys. Plasmas}, 17\penalty0 (1):\penalty0 012101, 2010.

\bibitem[F\"ul\"op and Moradi(2011)]{Fulop2011}
T.~F\"ul\"op and S.~Moradi.
\newblock Effect of poloidal asymmetry on the impurity density profile in
  tokamak plasmas.
\newblock \emph{Phys. Plasmas}, 18\penalty0 (3):\penalty0 030703, 2011.

\bibitem[Nordman et~al.(2008)Nordman, Singh, F\"ul\"op, Eriksson, Dumont,
  Andersson, Kaw, Strand, Tokar, and Weiland]{Nordman2008}
H.~Nordman, R.~Singh, T.~F\"ul\"op, L.-G. Eriksson, R.~Dumont, J.~Andersson,
  P.~Kaw, P.~Strand, M.~Tokar, and J.~Weiland.
\newblock Influence of the radio frequency ponderomotive force on anomalous
  impurity transport in tokamaks.
\newblock \emph{Phys. Plasmas}, 15:\penalty0 042316, 2008.

\bibitem[Angioni and Peeters(2006)]{Angioni2006}
C.~Angioni and A.~G. Peeters.
\newblock Direction of impurity pinch and auxiliary heating in tokamak plasmas.
\newblock \emph{Phys. Rev. Lett.}, 96:\penalty0 095003, 2006.

\bibitem[Nordman et~al.(2007)Nordman, F\"ul\"op, Candy, Strand, and
  Weiland]{Nordman2007a}
H.~Nordman, T.~F\"ul\"op, J.~Candy, P.~Strand, and J.~Weiland.
\newblock Influence of magnetic shear on impurity transport.
\newblock \emph{Phys. Plasmas}, 14\penalty0 (5):\penalty0 052303, 2007.

\bibitem[Angioni et~al.(2007)Angioni, Carraro, Dannert, Dubuit, Dux, Fuchs,
  Garbet, Garzotti, Giroud, Guirlet, Jenko, Kardaun, Lauro-Taroni, Mantica,
  Maslov, Naulin, Neu, Peeters, Pereverzev, Puiatti, P\"utterich, Stober,
  Valovi\v{c}, Valisa, Weisen, Zablotsky, {ASDEX Upgrade team}, and {JET--EFDA
  contributors}]{Angioni2007}
C.~Angioni, L.~Carraro, T.~Dannert, N.~Dubuit, R.~Dux, C.~Fuchs, X.~Garbet,
  L.~Garzotti, C.~Giroud, R.~Guirlet, F.~Jenko, O.~J. W.~F. Kardaun,
  L.~Lauro-Taroni, P.~Mantica, M.~Maslov, V.~Naulin, R.~Neu, A.~G. Peeters,
  G.~Pereverzev, M.~E. Puiatti, T.~P\"utterich, J.~Stober, M.~Valovi\v{c},
  M.~Valisa, H.~Weisen, A.~Zablotsky, {ASDEX Upgrade team}, and {JET--EFDA
  contributors}.
\newblock Particle and impurity transport in the {Axial Symmetric Divertor
  Experiment Upgrade} and the {Joint European Torus}, experimental observations
  and theoretical understanding.
\newblock \emph{Phys. Plasmas}, 14\penalty0 (5):\penalty0 055905, 2007.

\bibitem[Angioni et~al.(2009)Angioni, Peeters, Pereverzev, Bottino, Candy, Dux,
  Fable, Hein, and Waltz]{Angioni2009a}
C.~Angioni, A.~G. Peeters, G.~V. Pereverzev, A.~Bottino, J.~Candy, R.~Dux,
  E.~Fable, T.~Hein, and R.~E. Waltz.
\newblock Gyrokinetic simulations of impurity, {He} ash and $\alpha$ particle
  transport and consequences on {ITER} transport modelling.
\newblock \emph{Nucl. Fusion}, 49\penalty0 (5):\penalty0 055013, 2009.

\bibitem[F\"ul\"op and Nordman(2009)]{Fulop2009}
T.~F\"ul\"op and H.~Nordman.
\newblock Turbulent and neoclassical impurity transport in tokamak plasmas.
\newblock \emph{Phys. Plasmas}, 16\penalty0 (3):\penalty0 032306, 2009.

\bibitem[Nordman et~al.(2011)Nordman, Skyman, Strand, Giroud, Jenko, Merz,
  Naulin, Tala, and {the JET--EFDA contributors}]{Nordman2011}
H.~Nordman, A.~Skyman, P.~Strand, C.~Giroud, F.~Jenko, F.~Merz, V.~Naulin,
  T.~Tala, and {the JET--EFDA contributors}.
\newblock Fluid and gyrokinetic simulations of impurity transport at {JET}.
\newblock \emph{Plasma Phys. Contr. F.}, 53\penalty0 (10):\penalty0 105005,
  2011.

\bibitem[Skyman et~al.(2011)Skyman, Nordman, and Strand]{Skyman2011a}
A.~Skyman, H.~Nordman, and P.~Strand.
\newblock Impurity transport in temperature gradient driven turbulence.
\newblock Submitted to Phys. Plasmas, 2011.
\newblock URL \url{arXiv:1107.0880}.

\bibitem[Jenko et~al.(2000)Jenko, Dorland, Kotschenreuther, and
  Rogers]{Jenko2000}
F.~Jenko, W.~Dorland, M.~Kotschenreuther, and B.~N. Rogers.
\newblock Electron temperature gradient driven turbulence.
\newblock \emph{Phys. Plasmas}, 7\penalty0 (5):\penalty0 1904, 2000.

\bibitem[Dannert(2005)]{Dannert2005a}
T.~Dannert.
\newblock \emph{Gyrokinetische {S}imulation von {P}lasmaturbulenz mit
  gefangenen {T}eilchen und elektromagnetischen {E}ffekten}.
\newblock Ph.d. thesis (monography), Technischen {U}niversit\"at {M}\"unchen,
  2005.

\bibitem[Dannert and Jenko(2005)]{Dannert2005b}
T.~Dannert and F.~Jenko.
\newblock Gyrokinetic simulation of collisionless trapped-electron mode
  turbulence.
\newblock \emph{Phys. Plasmas}, 12\penalty0 (7):\penalty0 072309, 2005.

\bibitem[Weiland(2000)]{Weiland2000}
J.~Weiland.
\newblock \emph{Collective Modes in Inhomogeneous Plasmas}.
\newblock IoP Publishing, 2000.

\bibitem[Hirose et~al.(1994)Hirose, Zhang, and Elia]{Hirose1994}
A.~Hirose, L.~Zhang, and E.~Elia.
\newblock Higher order collisionless ballooning mode in tokamaks.
\newblock \emph{Phys. Rev. Lett.}, 72\penalty0 (25):\penalty0 3993--3996, 1994.

\bibitem[Lin()]{Lindgren}
Lindgren.
\newblock URL \url{http://www.pdc.kth.se/resources/computers/lindgren/}.

\bibitem[HPC()]{HPC-FF}
{HPC-FF}.
\newblock URL \url{http://www2.fz-juelich.de/jsc/juropa/}.

\end{thebibliography}

\clearpage

\begin{table}[\figplacing]
 \centering
 \caption[Parameters]{\small Parameters used in the gyrokinetic simulations, \Note{\dag} denotes scan parameters}
 \label{tab:parameters}
 \begin{tabular}{l||r|r}\hline
 & $R/L_{n_e}$-scaling: & $Z$-scaling: \\ \hline\hline
 $T_i/T_e$:             & $1.0$     & $1.0$     \\
 $s$:                   & $0.8$     & $0.8$     \\
 $q$:                   & $1.4$     & $1.4$     \\
 $\beta$:               & $10^{-4}$ & $10^{-4}$ \\
 $\epsilon=r/R$:        & $0.14$    & $0.14$    \\
 $n_e$, $n_i+n_Z$:      & $1.0$     & $1.0$     \\ 
 $n_Z$ \emph{(trace)}:  & $10^{-6}$ & $10^{-6}$ \\
 $R/L_{T_i},R/L_{T_Z}$:             & $2.0$         & $2.0$ \\
 $R/L_{T_e}$:                       & $5.0$         & $5.0$ \\ 
 $R/L_{n_{i,e}}$:\Note{\dag}        & $1.0$--$13.0$ & $5.0$--$13.0$ \\
 $Z$:\Note{\dag}                    & $2$, $28$     & $2$--$42$ 
 \end{tabular}
\end{table}

\clearpage


\begin{figure}[\figplacing]
 \centering
 \subfloat[time series and time averages of the main ion flux $\left(\Gamma_p\right)$ from NL~GENE simulations \label{fig:time_series}]
{\includegraphics[width=\figwidth]{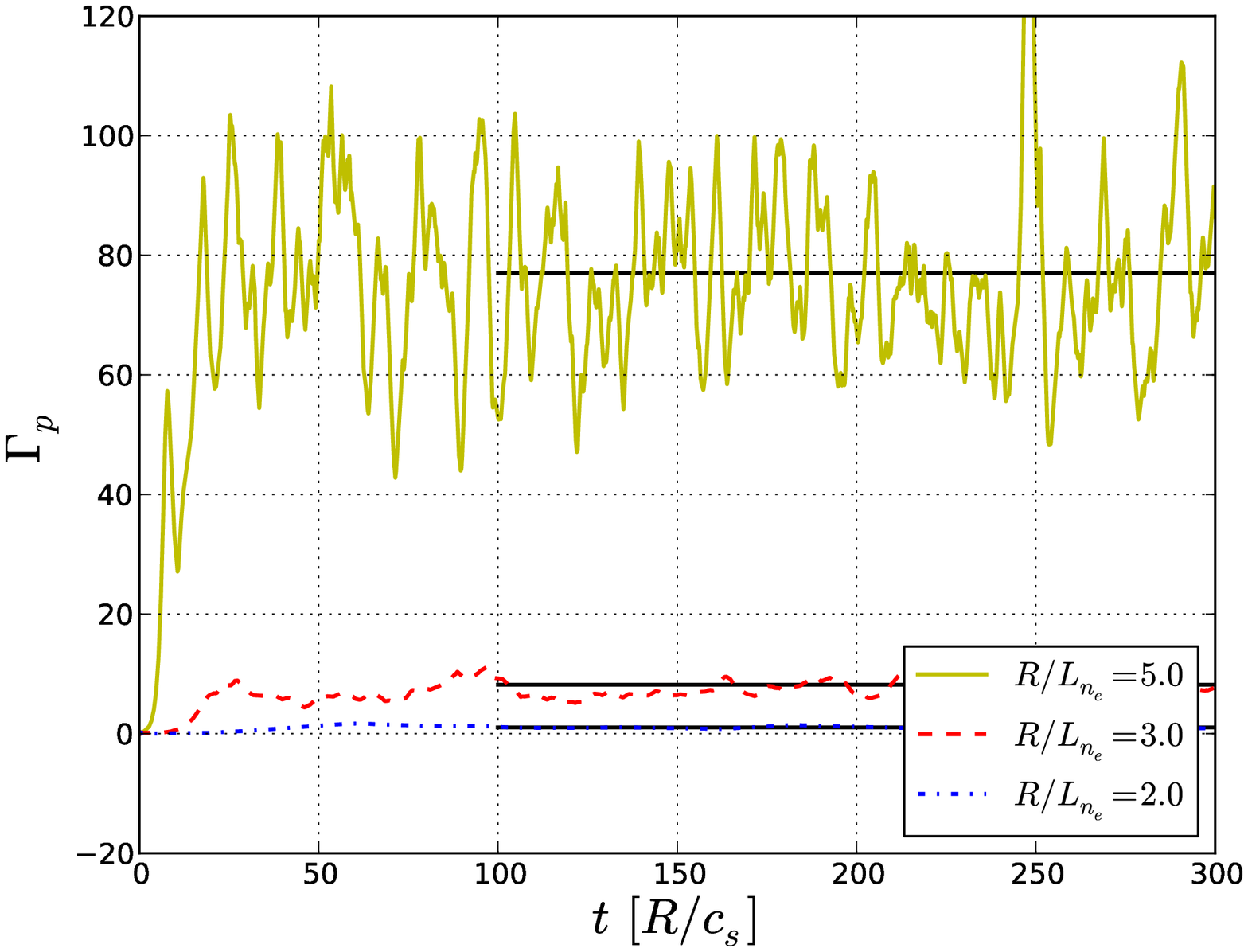}}
\phantomcaption{}
\end{figure}

\begin{figure}[\figplacing]
 \ContinuedFloat
 \centering
 \subfloat[main ion flux $\left(\Gamma_p\right)$ dependence on the background density gradient ($R/L_{n_e}$). 
\label{fig:Gamma_p}]
{\includegraphics[width=\figwidth]{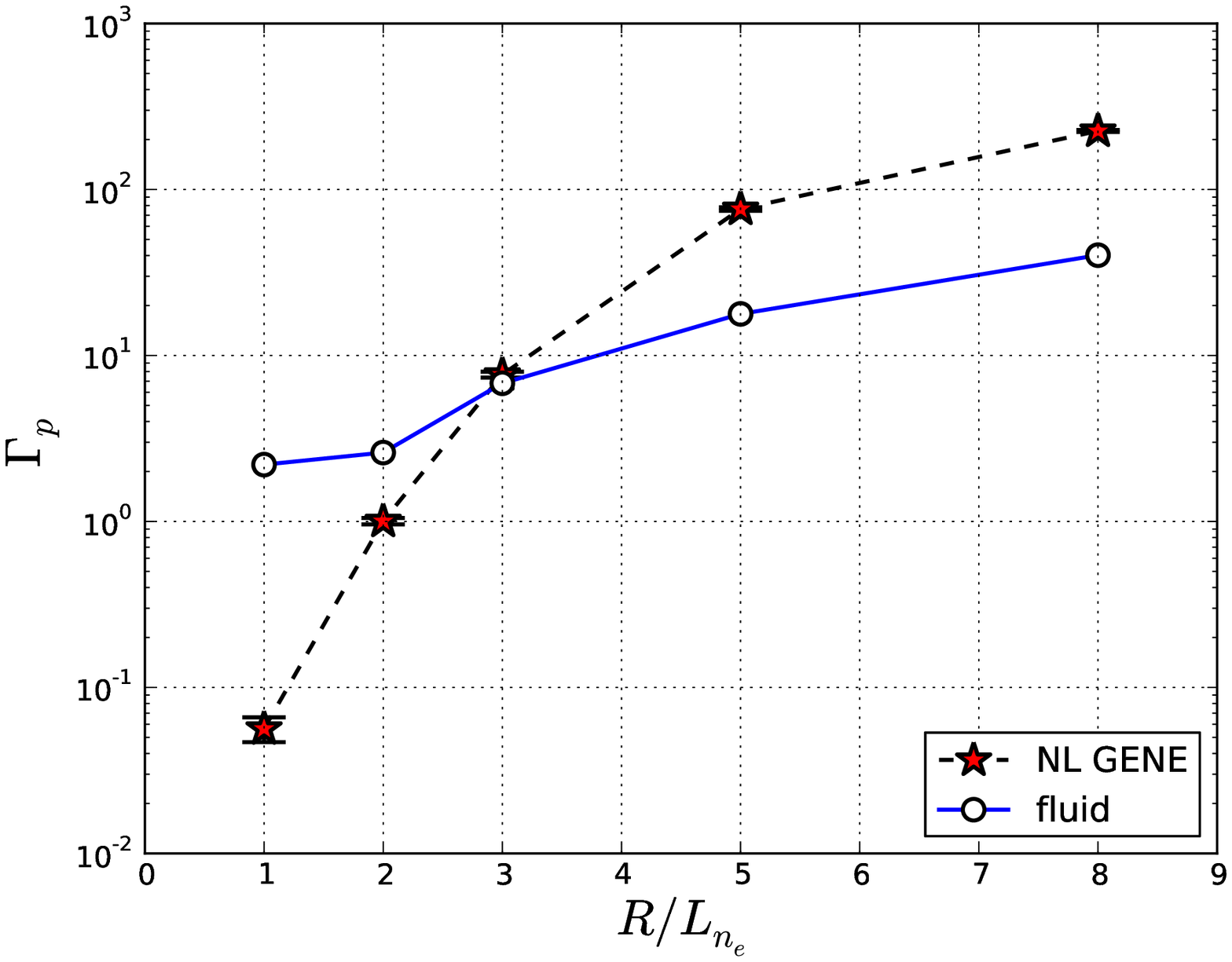}}
 \caption{Main ion flux $\left(\Gamma_p\right)$ dependence on the background electron density gradient ($-R\grad n_e/n_e=R/L_{n_e}$). 
NL~GENE and fluid data with protons as main ions. Parameters are $q=1.4$, $s=0.8$, $\epsilon=r/R=0.14$, $R/L_{T_i, Z}=2.0$, $R/L_{T_e}=5.0$, and $\tau=T_e/T_i=1.0$. 
The fluid data was obtained for $k_\theta \rho_s = 0.2$. The fluxes are normalised to $v_{T,i}n_e\rho_i^2/R^2$. 
The error bars indicate an estimated uncertainty of one standard deviation.}
\label{fig:main_ions}
\end{figure}

\clearpage

\begin{figure}[\figplacing]
 \centering
 \subfloat[dependence of the impurity peaking factor $\left(PF_Z\right)$ on the background density gradient \label{fig:omn_TEM}]
{\includegraphics[width=\figwidth]{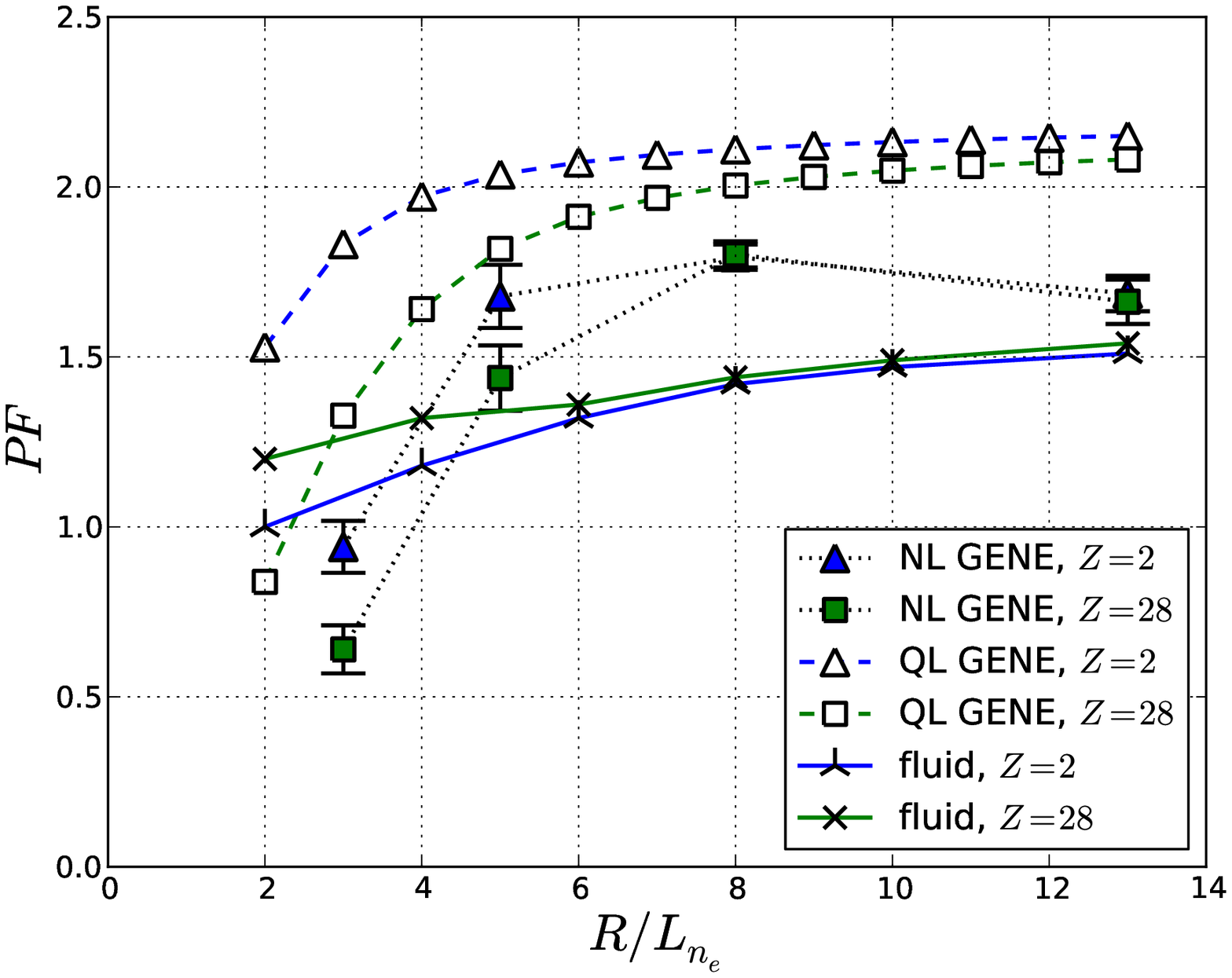}}
\phantomcaption{}
\end{figure}

\begin{figure}[\figplacing]
 \ContinuedFloat
 \centering
 \subfloat[dependence of the impurity diffusivity and convective velocity ($D_Z$ and $RV_Z$) on the background density gradient \label{fig:D_RV}]
{\includegraphics[width=\figwidth]{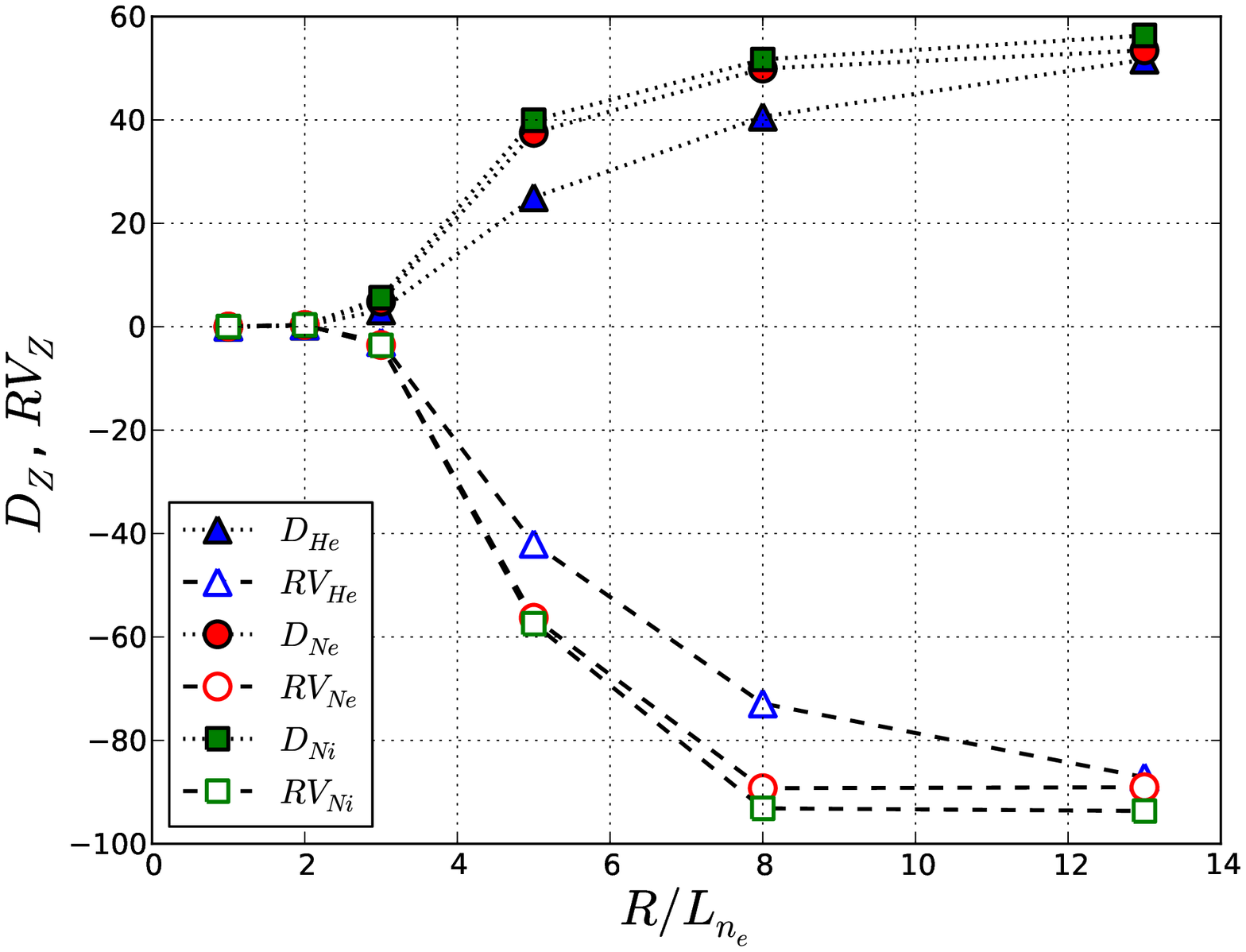}}
\phantomcaption{}
\end{figure}

\begin{figure}[\figplacing]
 \ContinuedFloat
 \centering
 \subfloat[scaling of real frequency ($\omega_r$) and growthrate ($\gamma$) with the background density gradient \label{fig:omn_TEM_eigens}]
{\includegraphics[width=\figwidth]{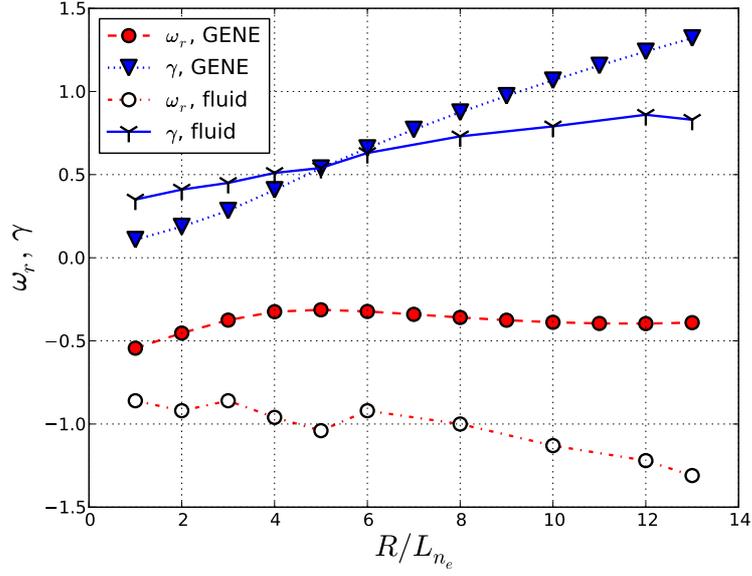}}
\caption{Scalings of the impurity peaking factor ($PF_Z=-RV_Z/D_Z$) with the background electron density gradient ($R/L_{n_e}$), with parameters as in figure~\ref{fig:main_ions}.
QL and fluid data have been acquired using $k_\theta\rho_s=0.2$.
Figure~\ref{fig:D_RV} shows the diffusivities and pinches corresponding to the NL~GENE impurity peaking factors ($PF_Z$) in figure~\ref{fig:omn_TEM}.
$D_Z$ and $RV_Z$ are normalised to $v_{T,i}\rho_i^2/R$.
The eigenvalues in figure~\ref{fig:omn_TEM_eigens} are from fluid and GENE simulations, and are normalised to $c_s/R$.
The error bars indicate an estimated uncertainty of one standard deviation.}
\label{fig:omn}
\end{figure}
\clearpage

\begin{figure}[\figplacing]
 \centering
 \includegraphics[width=\figwidth]{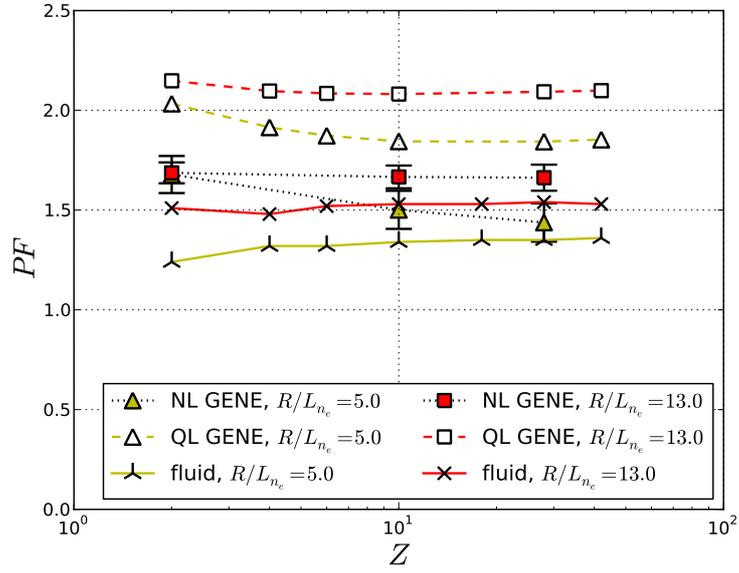}
 \caption{Scaling of the impurity peaking factor ($PF_Z=-RV_Z/D_Z$) with impurity charge $Z$, with parameters as in figure~\ref{fig:main_ions}; $k_\theta\rho_s=0.2$ was used in the QL and fluid simulations.
The error bars indicate an estimated uncertainty of one standard deviation.}
 \label{fig:Z}
\end{figure}
\clearpage

%
%

\end{document}